\documentclass[preprint,aps,showpacs,nofootinbib]{revtex4}

\usepackage{epsfig,amssymb,amsmath}

\def\bea{\begin{eqnarray}}
\def\eea{\end{eqnarray}}
\def\bec{\begin{center}}
\def\ec{\end{center}}

\def\beq{\begin{equation}}
\def\eeq{\end{equation}}

\begin{document}
\draft
\tighten
\title{\large \bf Probing the Messenger of SUSY Breaking with Gaugino Masses\footnote{Talk given by K. Choi at International Workshop on
Theoretical High Energy Physics, March 2007, Roorkee, India}}
\author{Won Sang Cho\footnote{email:wscho@hep.kaist.ac.kr} and
Kiwoon Choi\footnote{email: kchoi@hep.kaist.ac.kr}}
\address{
Department of Physics, KAIST, Daejeon 305-701, Korea}
\date{\today}

\vspace{1cm}

\begin{abstract}
Gaugino masses might provide useful information on the underlying
scheme of supersymmetry breaking as they are least dependent on the
unknown physics between  the TeV scale and the high messenger scale
of supersymmetry breaking.
 We discuss the pattern of low energy gaugino masses in various schemes of
supersymmetry breaking together with the possibility to determine
the gaugino masses at  LHC.

\end{abstract}
 \maketitle

\section{introduction}

Low energy supersymmetry (SUSY) \cite{nilles} is one of the prime
candidates for physics beyond the standard model at the TeV scale.
 Most phenomenological aspects of low energy SUSY are determined by
the soft  SUSY breaking terms in low energy effective lagrangian.
Those soft terms are generated at certain messenger scale $M_{\rm
mess}$ presumed to be higher than TeV, and then receive  quantum
corrections due to the renormalization group (RG) evolution and
threshold effects that might occur at scales below $M_{\rm mess}$.
Among all the soft terms, the MSSM gaugino masses $M_a$
$(a=SU(3),SU(2),U(1))$ appear to be the least model dependent as
they are related to the corresponding standard model (SM) gauge
coupling constants $g_a$ in a  nontrivial manner. Specifically,
$M_a/g_a^2$ do not run at the one-loop level, and also
possible intermediate threshold corrections to $M_a/g_a^2$ are
severely constrained if one requires to keep the successful gauge
coupling unification at $M_{GUT}\sim 2\times 10^{16}$ GeV. In this
respect, analysis of the gaugino mass pattern at TeV can be
considered as a promising first step to uncover the mediation
mechanism of SUSY breaking at $M_{\rm mess}$. In this talk, we
discuss the possible pattern of low energy gaugino masses which
might be obtained in various SUSY breaking schemes \cite{choinilles}
and also the possibility to determine the gaugino masses at  LHC
\cite{cascade,cho,cho1}, aiming to see what kind of information on
SUSY breaking scheme can be extracted once the low energy gaugino
masses can be determined by future collider experiments.

 \section{generic gaugino masses in 4d supergravity}

In  4D effective supergravity (SUGRA) with the cutoff scale
$\Lambda$ which is chosen to be just below the string or
Kaluza-Klein (KK) or GUT threshold scale, the running gauge
couplings and gaugino masses at a scale $\mu$ below $\Lambda$ (but
above the next threshold scale $M_{\rm th}$) are determined by the
gauge coupling superfield ${\cal F}_a(p^2)$ ($M^2_{\rm
th}<p^2<\Lambda^2$)  in the quantum effective action of gauge
superfields: \bea \Gamma=\int d^4p\, d^4\theta \left(\,
\frac{1}{4}{\cal F}_a(p^2) W^{a\alpha}\frac{{\cal D}^\alpha{\cal
D}_\alpha}{16 p^2}W^a_\alpha+{\rm h.c}\,\right),\eea where
$W^a_\alpha$ denote the chiral gauge superfields and ${\cal
D}_\alpha$ is super-covariant derivative. At one-loop approximation,
${\cal F}_a$ is given by \cite{kaplunovsky} \bea
\label{gaugecouplingsuperfield}{\cal F}_a(p^2)&=&{\rm
Re}(f^{(0)}_a)-\frac{1}{16\pi^2}(3C_a-\sum_iC_a^i)\ln\left(\frac{CC^*\Lambda^2}{p^2}\right)
\nonumber \\&&-\frac{1}{8\pi^2}\sum_i
C_a^i\ln\left(e^{-K_0/3}Z_i\right)+\frac{1}{8\pi^2}{\Omega}_a,\eea
where $f_a^{(0)}$ are the tree-level holomorphic gauge kinetic
function,  $C_a$ and $C_a^i$ are the quadratic Casimir of the gauge
multiplet and the gauge-charged matter superfield  $Q_i$,
respectively, and $C$ is the chiral compensator of 4D SUGRA. Here
$K_0(X_I,X_I^*)$ is the K\"ahler potential of generic SUSY breaking
(moduli or matter) superfields $X_I$ which have nonzero
$F$-components $F^I$, $Z_i(X_I,X_I^*)$ is the K\"ahler metric of
$Q_i$, and $\Omega_a$ include the string, KK and GUT threshold
corrections as well as the (regularization scheme-dependent)
field-theoretic one-loop part: $\frac{1}{8\pi^2}C_a\ln[{\rm
Re}(f_a^{(0)})]$.
In the one-loop approximation, $\Omega_a$ are independent of the
external momentum $p^2$, thus independent of $C$ as a consequence of
the super-Weyl invariance. However $\Omega_a$ generically depend on
SUSY breaking  fields $X_I$, and a full determination of their
$X_I$-dependence requires a detailed knowledge of the UV physics
above $\Lambda$. From the above  gauge coupling superfield, one
easily finds that the running gauge couplings and gaugino masses at
$\mu$ ($M_{\rm th}<\mu<\Lambda$) are given by
\cite{kaplunovsky,anomaly,choinilles}
 \bea
\frac{1}{g_a^2(\mu)}&=&{\cal F}_a|_{C=e^{K_0/6},\,p^2=\mu^2}
\nonumber \\
&=& {\rm Re}(f^{(0)}_a) -\frac{1}{16\pi^2}\left[(3C_a-\sum_i
C_a^i)\ln\left(\frac{\Lambda^2}{\mu^2}\right)\right.\nonumber \\
&&+\left.(C_a-\sum_iC_a^i)K_0+2\sum_iC_a^i\ln
Z_i\,\right]+\frac{1}{8\pi^2}\Omega_a,\nonumber \\
\frac{M_a(\mu)}{g_a^2(\mu)}&=& F^A\partial_A{\cal
F}_a|_{{C=e^{K_0/6},\,p^2=\mu^2}}\nonumber \\
&=& F^I\left[\frac{1}{2}\partial_I f_a^{(0)}-\frac{1}{8\pi^2}\sum_i
C_a^i\partial_I\ln(e^{-K_0/3}Z_i)+\frac{1}{8\pi^2}\partial_I\Omega_a\right]
\nonumber
\\
&&-\,\frac{1}{16\pi^2}(3C_a-\sum_iC_a^i)\frac{F^C}{C},\eea where
$F^A=(F^C,F^I)$, $\partial_A=(\partial_C,\partial_I)$,
$\frac{F^C}{C}=m^*_{3/2}+\frac{1}{3}F^I\partial_IK_0$, and
$C=e^{K_0/6}$ corresponds to the Einstein frame condition.

The above expression of $M_a/g_a^2$ is valid at any scale between
$M_{\rm th}$ and $\Lambda$. However, depending upon  the SUSY
breaking scenario, $M_a/g_a^2$ can receive important threshold
correction at the next threshold scale $M_{\rm th}$.
 To see how  $M_a/g_a^2$ are modified by threshold effect,
let us assume $\{Q_i\}\equiv \{\Phi+\Phi^c,Q_x\}$ and $\Phi+\Phi^c$
get a supersymmetric mass of the order of $M_{\rm th}$,  while $Q_x$
remain to be massless at $M_{\rm th}$. Then $\Phi+\Phi^c$ can be
integrated out to derive the low energy parameters at scales below
$M_{\rm th}$. The relevant couplings of $\Phi+\Phi^c$ at $M_{\rm
th}$ can be written as \bea \int d^4\theta
CC^*e^{-K_0/3}\left(Z_\Phi\Phi^*\Phi+Z_{\Phi^c}\Phi^{c*}\Phi^c\right)
+\left(\int d^2\theta \,C^3\lambda_\Phi X_\Phi\Phi^c\Phi+{\rm
h.c}\right), \eea where $X_\Phi$ is assumed to have a vacuum value
$\langle X_\Phi\rangle= {M}_\Phi +\theta^2 F^{X_\Phi}$ for which the
physical mass of $\Phi+\Phi^c$ is given by ${\cal
M}_\Phi=\lambda_\Phi{C X_\Phi}/{\sqrt{e^{-2K_0/3}Z_\Phi
Z_{\Phi^c}}}$.  Then, integrating out $\Phi+\Phi^c$ yields a
threshold correction to ${\cal F}_a$:
 \bea
 \Delta {\cal F}_a(M_{\rm th})=
- \frac{1}{8\pi^2}\sum_{\Phi} C_a^\Phi \ln\left(\frac{{\cal M}_\Phi
 {\cal M}_\Phi^*}{M_{\rm th}^2}\right),
 \eea
which results in the threshold correction to gaugino mass at $M_{\rm
th}$: \bea &&M_a(M_{\rm th}^-)-M_a(M_{\rm th}^+)= g_a^2(M_{\rm
th})F^A\partial_A \Delta {\cal F}_a \nonumber
\\&=&-\frac{g_a^2(M_{\rm th})}{8\pi^2} \sum_\Phi C_a^\Phi
\left(\frac{F^C}{C}+\frac{F^{X_\Phi}}{M_\Phi}-
F^I\partial_I\ln(e^{-2K_0/3}Z_\Phi Z_{\Phi^c})\right). \eea
Including this threshold, one finds
 \bea \left(\frac{M_a}{g_a^2}\right)_{M_{\rm th}^-}&=&
F^I\left[\frac{1}{2}\partial_I f_a^{(0)} -\frac{1}{8\pi^2}\sum_x
C_a^x\partial_I\left(e^{-K_0/3}Z_x\right)+\frac{1}{8\pi^2}\partial_I\Omega_a\right]\nonumber
\\
&&-\frac{1}{8\pi^2}\sum_\Phi C_a^\Phi
\frac{F^{X_\Phi}}{M_\Phi}-\frac{1}{16\pi^2}(3C_a-\sum_xC_a^x)\frac{F^C}{C},
\eea where $\sum_x$ denotes the summation over $\{ Q_x\}$ which
remain as light matter fields at $M_{\rm th}^-$.

 One can repeat the above procedure, i.e. run down to the
lower threshold scale, integrate out the massive fields there, and
then include the threshold correction to gaugino masses until one
arrives at the TeV scale. Then one finally finds \cite{choinilles}
\bea\label{lowscaleratio} \left(\frac{M_a}{g_a^2}\right)_{\rm TeV}
=\tilde{M}_a^{\rm moduli}+ \tilde{M}_a^{\rm gauge}+\tilde{M}_a^{\rm
conformal}+\tilde{M}_a^{\rm konishi}+\tilde{M}_a^{\rm UV} \eea
 where \bea\label{components}
\tilde{M}_a^{\rm moduli}&=&\frac{1}{2}F^I\partial_If_a^{(0)},
\nonumber \\\tilde{M}_a^{\rm gauge}&=& -\frac{1}{8\pi^2}\sum_\Phi
C_a^\Phi\frac{F^{X_\Phi}}{M_\Phi}, \nonumber \\
\tilde{M}_a^{\rm
conformal}&=&\frac{1}{16\pi^2}b_a\frac{F^C}{C},\nonumber \\
\tilde{M}_a^{\rm konishi}
&=&-\frac{1}{8\pi^2}\sum_mC_a^mF^I\partial_I\ln(e^{-K_0/3}Z_m),
\nonumber \\
\tilde{M}_a^{\rm UV}&=&\frac{1}{8\pi^2}F^I\partial_I \Omega_a,\eea
where $\sum_m$ denotes the summation over the light matter
multiplets $\{Q_m\}$ at the TeV scale, $\sum_\Phi$ denotes the
summation over the gauge messenger fields $\Phi+\Phi^c$ which have a
mass lighter than $\Lambda$ but heavier than TeV, and
$b_a=-3C_a+\sum_m C_a^m$ are the one-loop beta-function coefficients
at TeV. Here $\tilde{M}_a^{\rm moduli}$ denotes the moduli-mediated
tree level value of $M_a/g_a^2$ \cite{moduli}, $\tilde{M}_a^{\rm
gauge}$  is the intermediate scale gauge threshold due to
gauge-charged massive particles with a mass between $\Lambda$ and
TeV \cite{gauge},  $\tilde{M}_a^{\rm conformal}$ is the
anomaly-mediated contribution determined by the conformal anomaly at
TeV \cite{anomaly}, $\tilde{M}_a^{\rm konishi}$ is a piece
determined by the Konishi anomaly \cite{konishi},
and finally $\tilde{M}_a^{\rm UV}$ contains the {\it UV thresholds}
at string, KK and GUT scales.

Formulae (\ref{lowscaleratio}) and (\ref{components}) give the most
general description of gaugino masses and its origin from the
underlying schemes \cite{choinilles}. Depending upon the SUSY
breaking scenario, $M_a/g_a^2$ are dominated by some of these five
contributions. The SM gauge coupling constants at TeV have been
measured with the (approximate) result: $g_1^2:g_2^2:g_3^2\simeq
1:2:6$.  As a result, once the gaugino mass ratios at TeV are
measured, the ratios of $M_a/g_a^2$ at TeV can be experimentally
determined, which will allow us to test SUSY breaking schemes using
the predicted pattern of low energy gaugino masses.

As $f_a^{(0)}$ determine the gauge coupling constants at $M_{GUT}$,
it is expected that  $\tilde{M}_a^{\rm moduli}$ are {\it universal}
in most cases realizing the gauge coupling unification at $M_{GUT}$.
In compactified string theory or higher dimensional SUGRA,  the tree
level gauge kinetic functions are generically given by
$f_a^{(0)}=\sum_Ik_{aI}X_I$, where $X_I$ correspond to the dilaton
and/or moduli superfields and $k_{aI}$ are rational numbers. In
models realizing gauge coupling unification, $k_{aI}$ are universal
for the SM gauge group $a=SU(3),SU(2),U(1)$, which would give
universal $\tilde{M}_a^{\rm moduli}=\frac{1}{2}\sum_Ik_{aI}F^I$.

The intermediate gauge thresholds $\tilde{M}_a^{\rm gauge}=
-\frac{1}{8\pi^2}\sum_\Phi C_a^\Phi\frac{F^{X_\Phi}}{M_\Phi}$
accompany the additional running of gauge couplings from $M_{GUT}$
to $M_\Phi$:  $\Delta ({1}/{g_a}^2)=\frac{1}{4\pi^2}\sum_\Phi
c_a^\Phi\ln(M_{GUT}/M_\Phi)$, indicating that  $\tilde{M}_a^{\rm
gauge}$ with $M_\Phi \ll M_{GUT}$ are required to be universal also
to keep the gauge coupling unification at $M_{GUT}$. On the other
hand, there is {\it no} good reason to expect that the string, KK
and GUT thresholds $\tilde{M}_a^{\rm UV}$ are universal.
 In fact,  the  UV thresholds encoded in
$\frac{1}{8\pi^2}\Omega_a$ are most model-dependent, and difficult
to compute.  If this part gives an important contribution to
$M_a/g_a^2$, it is difficult to make a model-independent statement
about the gaugino masses.

With the above observation, one can consider  the following three
distinctive  patterns of low energy gaugino masses which can result
from theoretically well motivated setup.

\vskip 0.2cm
 \noindent {\bf mSUGRA pattern}: The scenario which has been discussed most often in the
literatures  is that $(M_a/g_a^2)_{\rm TeV}$ are dominated by
$\tilde{M}_a^{\rm moduli}$ or $\tilde{M}_a^{\rm gauge}$ which are
assumed to be {\it universal}: \bea
\left(\frac{M_a}{g_a^2}\right)_{\rm TeV}\,\simeq\,\,
\tilde{M}_a^{\rm moduli}\,\,\,\, \mbox{or}\, \,\,\,\tilde{M}_a^{\rm
gauge},\eea leading to the following low energy gaugino mass ratios:
\bea M_1:M_2:M_3 \simeq 1 : 2:6 \eea which will be termed mSUGRA
pattern in the following. Schemes giving the mSUGRA pattern of
gaugino masses include the dilaton and/or moduli dominated SUSY
breaking scenarios realized in various compactified string theories
\cite{moduli} with a large string and compactification scales near
$M_{GUT}\sim 2\times 10^{16}$ GeV, gaugino mediation scenario
\cite{gaugino}, and also gauge mediation scenario \cite{gauge} with
a messenger scale $M_{\rm mess}\ll M_{GUT}$.

We stress that the universality of $\tilde{M}_a^{\rm moduli}$ which
is essential for the mSUGRA pattern heavily relies on the assumption
of high scale gauge coupling unification. In models without gauge
coupling unification, $k_{aI}$ and thus $\tilde{M}_a^{\rm moduli}$
are generically non-universal and highly model-dependent. However,
in some case, one might  be able to extract  information on $k_{aI}$
for $X_I$ providing a dominant source of SUSY breaking, thereby make
a certain prediction on low energy gaugino masses. A nontrivial
example of this kind is the large volume compactification of Type
IIB string theory proposed in \cite{conlon}. In the model of
\cite{conlon}, moduli are stabilized at a vacuum
with the string scale $M_{st}\sim 10^{11}$ GeV, and
$f_a^{(0)}=k_aT_s+h_aS$, where $T_s$ is the volume modulus of small
4-cycle  and $S$ is the IIB dilaton with $|F^S|\ll |F^{T_s}|$. As
${\rm Re}(f_a^{(0)})\simeq 1/g_a^2(M_{st})$, in such intermediate
string scale scenario, $k_a$ and $h_a$ can not be constrained by
gauge coupling unification.
  However, the model of
\cite{conlon} has $k_{SU(3)}=k_{SU(2)}$ regardless of the values of
$g_a^2(M_{st})$, while $k_{U(1)}$ and $h_a$ are generically
non-universal independent parameters. As $F^S$ and $F^C$ ($C=$ SUGRA
compensator) are negligible in the model of \cite{conlon},
$k_{SU(3)}=k_{SU(2)}$ leads to $M_2:M_3\,\simeq\, 1:3$, while the
ratio with $M_1$ depends on the unknown $k_{U(1)}/k_{SU(2)}$.

\vskip 0.2cm \noindent {\bf Mirage pattern}: Another interesting
scenario is that $(M_a/g_a^2)_{\rm TeV}$ are dominated by
$\tilde{M}_a^{\rm conformal}$ and universal $\tilde{M}_a^{\rm
moduli}$ (or $\tilde{M}_a^{\rm gauge}$) which are comparable to each
other: \bea \left(\frac{M_a}{g_a^2}\right)_{\rm TeV}\,\simeq\,
\left(\tilde{M}_a^{\rm moduli}\,\,\,\, \mbox{or}\,
\,\,\,\tilde{M}_a^{\rm gauge}\right)+\tilde{M}_a^{\rm
conformal},\eea leading to \bea M_1:M_2:M_3 \simeq (1+0.66\alpha) :
(2+0.2\alpha):(6-1.8\alpha), \eea where $\alpha$ is {\it a positive
parameter of order unity} defined as \bea
\frac{g_{GUT}^2}{16\pi^2}b_a\alpha\ln(M_{Pl}/m_{3/2}) \,\equiv\,\,
\frac{\tilde{M}_a^{\rm conformal}}{\tilde{M}_a^{\rm moduli}} \,\,\,
\mbox{or}\,\,\, \frac{\tilde{M}_a^{\rm conformal}}{\tilde{M}_a^{\rm
gauge}},\eea where $b_a=(\frac{33}{5},1,-3)$ are the MSSM beta
function coefficients. This pattern is termed mirage pattern  as
$M_a$ are unified at the mirage messenger scale \cite{mirage1}: \bea
M_{\rm mirage}=M_{GUT}(m_{3/2}/M_{Pl})^{\alpha/2}.\eea Examples
giving the mirage pattern of gaugino masses include the KKLT
compactification \cite{kklt} of Type IIB string theory with the MSSM
gauge fields living on $D7$ branes \cite{mirage2}, deflected anomaly
mediation scenario proposed in \cite{dam}, and also some variants of
KKLT setup \cite{variant}.

Mirage pattern might be considered as a smooth interpolation between
the mSUGRA pattern ($\alpha=0$) and the anomaly pattern
($\alpha=\infty$). For a positive $\alpha={\cal O}(1)$ which is
predicted to be the case in most of the schemes yielding the mirage
pattern, gaugino masses  are significantly more degenerate than
those in mSUGRA and anomaly patterns. Different schemes giving the
same mirage pattern of gaugino masses can be distinguished from each
other by sfermion masses. For instance, in mirage mediation scheme
\cite{mirage1,mirage2} resulting from KKLT-type string
compactification \cite{kklt}, the 1st and 2nd generations of squark
and slepton masses   show up the same mirage unification at $M_{\rm
mirage}$, while the sfermion masses in deflected anomaly mediation
scenario have a  different structure \cite{dam}.

 \vskip
0.2cm \noindent {\bf Anomaly pattern}: If there is no singlet with
nonzero $F$-component or all SUSY breaking fields are sequestered
from the visible gauge fields,  $(M_a/g_a^2)_{\rm TeV}$ are
dominated by $\tilde{M}_a^{\rm conformal}$ \cite{anomaly}:\bea
\left(\frac{M_a}{g_a^2}\right)_{\rm TeV}\,\simeq\, \tilde{M}_a^{\rm
conformal},\eea leading to \bea M_1:M_2:M_3 \simeq 3.3 : 1 : 9 \eea
which is termed   anomaly pattern. One stringy example giving the
anomaly pattern would be the KKLT compactification with the MSSM
gauge fields on $D3$ branes \cite{mirage2}.

We finally note that there are schemes in which $(M_a/g_a^2)_{\rm
TeV}$ receive an important contribution from the UV thresholds
$\tilde{M}_a^{\rm UV}$ at string or GUT scale \cite{acharya}.
Gaugino masses in such scheme are the most model-dependent, and one
needs to know the details of the model around the string  or GUT
scale in order to determine the low energy gaugino masses.

 \section{measuring gaugino masses at lhc}

For  $R$-parity conserving SUSY model with neutralino LSP, if gluino
or squarks are light enough to be copiously produced at LHC, some
superparticle masses can be experimentally determined by analyzing
the various invariant mass distributions for the decay products of
the gluino or squark decays \cite{cascade,cho}. The three gaugino
mass patterns discussed in the previous section can be clearly
distinguished by their prediction of the gluino to LSP neutralino
mass ratio: ${m_{\tilde{g}}}/{m_{\chi_1}}\gtrsim 6$ for mSUGRA
pattern and ${m_{\tilde{g}}}/{m_{\chi_1}}\gtrsim 9$, while
${m_{\tilde{g}}}/{m_{\chi_1}}$ can be significantly smaller than 6
in mirage pattern. (Note that a nonzero Higgsino component in the
LSP $\chi_1^0$ makes the LSP mass $m_{\chi_1}$ smaller than the
smallest gaugino mass.)

There are several LHC observables providing information on gaugino
masses, which are expected to be available under a mild assumption
on SUSY spectra. Let us suppose that the gluino has  a mass lighter
than 2 TeV and $\chi_1^0$ has a sizable Bino or Wino component, so
that there will be a copious production of gluino pairs at LHC,
subsequently decaying into four jets and two LSPs: ${\tilde
g}{\tilde g} \rightarrow q_1 q_2 {\tilde\chi}_1^0 q_3
 q_4 {\tilde\chi}_1^0.$
One observable useful for the determination of
$\{m_{\tilde{g}},m_{\chi_1}\}$  is the $M_{T2}$ variable
\cite{mt2,cho1} of this gluino pair decay:
\begin{eqnarray}
\label{mt2_1} M_{T2}^2(\tilde{g}\rightarrow qq\chi_1) &\equiv&
\min_{{\bf p}_{T1}^\chi+{\bf p}_{T2}^\chi={\bf p}_T^{\rm miss}}
\left[ {\rm max} \{ m_T^2 ({\rm {\bf p}}_T^{q_1},{\rm {\bf
p}}_T^{q_2},{\rm {\bf p}}_{T1}^\chi), m_T^{2}({\rm {\bf
p}}_T^{q_3},{\rm {\bf p}}_T^{q_4},{\rm {\bf p}}_{T2}^\chi) \}
\right],\eea where \begin{eqnarray} m_{T}^2({\bf p}_T^{q_1},{\bf
p}_T^{q_2},{\bf p}_T^\chi) &\equiv&  m_{\chi_1}^2 + 2 (E_T^{q_1}
E_T^\chi - {\bf p}_T^{q_1} \cdot {\bf p}_T^\chi)\nonumber\\&&+\, 2
(E_T^{q_2} E_T^\chi - {\bf p}_T^{q_2} \cdot {\bf p}_T^\chi)+ 2
(E_T^{q_1} E_T^{q_2} - {\bf p}_T^{q_1} \cdot {\bf p}_T^{q_2}) \eea
for $E_T=\sqrt{|{\bf p}_T|^2 + m^2}$. Here, ${\bf p}_T^{q_1}$ and
${\bf p}_T^{q_2}$ denote the transverse momentum of the quark jets
from the one gluino decay, ${\bf p}_T^{\rm miss}$ is the observed
missing transverse momentum, and we have ignored the light quark
masses. If squark also has a mass comparable to the gluino mass, so
that squark pairs can be copiously produced, the $M_{T2}$ variable
for the squark pair decay $\tilde{q}\tilde{q}\rightarrow
q_1\chi_1q_2\chi_1$ provides an information on
$\{m_{\tilde{q}},m_{\chi_1}\}$:
\begin{eqnarray}
\label{mt2_2} M_{T2}^2(\tilde{q}\rightarrow q\chi_1) &\equiv
&\min_{{\bf p}_{T1}^\chi+{\bf p}_{T2}^\chi={\bf p}_T^{\rm miss}}
\left[ {\rm max} \{ m_T^2 ({\rm {\bf p}}_T^{q_1},{\rm {\bf
p}}_{T1}^\chi), m_T^{2}({\rm {\bf p}}_T^{q_2},{\rm {\bf
p}}_{T2}^\chi) \} \right],\eea where
\begin{eqnarray}
m_{T}^2 ({\bf p}_T^q,{\bf p}_T^\chi) \equiv  m_{\chi}^2 + 2 (E_T^q
E_T^\chi - {\bf p}_T^q \cdot {\bf p}_T^\chi).
\end{eqnarray}

The above two $M_{T2}$ variables will be available at LHC as long as
both the gluino and squark pairs are copiously produced at LHC, and
a sizable fraction of them decay into the LSP pair plus quark jets.
One still needs further information to determine the gluino to LSP
mass ratio. Such additional information can be provided by measuring
the maximal invariant mass $M^{\rm max}_{qq}$ of two jets in the
final products of the gluno decay $\tilde{g}\rightarrow
\tilde{q}q\rightarrow \chi_1 qq$. The observed two $M_{T2}$
variables (\ref{mt2_1}) and (\ref{mt2_2}) will tell us which of the
gluino and squark  is
heavier than the other. 
Then, under an appropriate event selection cut, the measured $M^{\rm
max}_{qq}$ corresponds to \bea M^{\rm max}_{qq}&\simeq&
\left[\frac{(m_{\tilde{g}}^2-m_{\tilde{q}}^2)(m_{\tilde{q}}^2-m_{\chi_1}^2)}{m_{\tilde{q}}^2}\right]^{1/2}
\quad \mbox{for heavier gluino}\nonumber \\
&&\mbox{or}\quad m_{\tilde{g}}-m_{\chi_1} \quad\quad \mbox{for
heavier squark}.\eea Combining this information with those from the
two $M_{T2}$ variables (\ref{mt2_1}) and (\ref{mt2_2}), one can
determine $\{m_{\tilde{g}},m_{\tilde{q}},m_{\chi_1}\}$ and thus the
gluino to LSP mass ratio.

In regard to the mass of the second lightest neutralino $\chi_2^0$,
a particularly interesting possibility is that $\chi_2^0$ is heavier
than slepton, e.g. $m_{\chi_2}-m_{\tilde{l}}\gtrsim 10$ GeV for
which the lepton from $\chi_2^0\rightarrow \tilde{l}l$ is energetic
enough to pass the selection cut, so  the following cascade decay of
squark is available \cite{cascade}: \bea \tilde{q}\rightarrow
q\chi_2^0\rightarrow q \tilde{l}^{\pm}l^{\mp} \rightarrow q \chi_1^0
l^+ l^-.\eea In such case, one can look at the edges of the
invariant mass distributions of $ll, llq$ and $lq$ to determine
$m_{\chi_2}$ and $m_{\tilde{l}}$. If $\chi_2^0$ is also
gaugino-like, the obtained value of $m_{\chi_2}$ will allow the full
determination of the gaugino mass ratios. Even when the slepton is
heavier than $\chi_2^0$, so the decay $\chi_2\rightarrow \tilde{l}l$
is not open, one can still determine $m_{\chi_2}-m_{\chi_1}$ using
the dilepton invariant mass distribution in the 3-body decay
$\chi_2\rightarrow \chi_1l^+l^-$.

\vspace{5mm} \noindent{\large\bf Acknowledgments} \vspace{5mm}

This work is supported by the KRF Grant funded by the Korean
Government (KRF-2005-201-C00006), the KOSEF Grant (KOSEF
R01-2005-000-10404-0), and the Center for High Energy Physics of
Kyungpook National University. We thank   K. S. Jeong and H. P.
Nilles for  useful discussions.



\bibliographystyle{aipproc}   


\IfFileExists{\jobname.bbl}{}
 {\typeout{}
  \typeout{******************************************}
  \typeout{** Please run "bibtex \jobname" to optain}
  \typeout{** the bibliography and then re-run LaTeX}
  \typeout{** twice to fix the references!}
  \typeout{******************************************}
  \typeout{}
 }

\end{document}